\documentclass[prl,onecolumn,groupedaddress,nofootinbib]{revtex4}
\usepackage{caption,subcaption,graphicx}
\usepackage[breaklinks=true,colorlinks=true,linkcolor=blue,urlcolor=blue,citecolor=blue]{hyperref}
\usepackage{mathrsfs}
\usepackage{relsize}
\usepackage{amsmath}
\usepackage{cleveref}
\usepackage{array,multirow}

\begin{document}

\title{Improving the Robustness of the Advanced LIGO Detectors to Earthquakes}
\author{E~Schwartz$^{1,4}$, A~Pele$^{1}$, J~Warner$^{2}$, B~Lantz$^{3}$, J~Betzwieser$^{1}$, 
K~L~Dooley$^{4,5}$,  
S~Biscans$^{6,15}$,  
M~Coughlin$^{7}$,
N~Mukund$^{8}$,
R~Abbott$^{6}$
% ...
}
\par\medskip
\address {$^{1}$LIGO Livingston Observatory, Livingston, LA 70754, USA }
\address {$^{2}$LIGO Hanford Observatory, Richland, WA 99352, USA }
\address {$^{3}$Stanford University, Stanford, CA 94305, USA }
\address {$^{4}$The University of Mississippi, University, MS 38677, USA }
\address {$^{5}$Cardiff University, Cardiff CF24 3AA, UK }
\address {$^{6}$LIGO, California Institute of Technology, Pasadena, CA 91125, USA }
\address {$^{7}$University of Minnesota, Minneapolis, MN 55455, USA }
\address {$^{8}$Max Planck Institute for Gravitational Physics (Albert Einstein Institute), D-30167 Hannover, Germany }
% ...
\address {$^{15}$LIGO, Massachusetts Institute of Technology, Cambridge, MA 02139, USA }
% ...

\begin{abstract}
Teleseismic, or distant, earthquakes regularly disrupt the operation of ground--based gravitational wave detectors such as Advanced LIGO. Here, we present \emph{EQ mode}, a new global control scheme, consisting of an automated sequence of optimized control filters that reduces and coordinates the motion of the seismic isolation platforms during earthquakes. This, in turn, suppresses the differential motion of the interferometer arms with respect to one another, resulting in a reduction of DARM signal at frequencies below 100\,mHz. Our method greatly improved the interferometers' capability to remain operational during earthquakes, with ground velocities up to 3.9\,$\mu \mbox{m/s}$ rms in the beam direction, setting a new record for both detectors. This sets a milestone in seismic controls of the Advanced LIGO detectors' ability to manage high ground motion induced by earthquakes, opening a path for further robust operation in other extreme environmental conditions.
\end{abstract}

\maketitle

\vspace{2pc}

\section{Introduction}
The detection of gravitational waves (GWs) by the Advanced Laser Interferometer Gravitational Wave Observatory (LIGO), include dozens of detections in recent years from binary black hole and neutron star mergers \cite{Abbott2016,Abbott2019,Abbott2017}, which were facilitated by the use of unparalleled seismic isolation systems. At their core, the GW detectors are modified km-scale Michelson interferometers with Fabry-Perot cavities in each arm, whose mirrors must approximate inertial test masses in order to act as markers of space-time coordinates. A passing gravitational wave induces a strain in space-time, which modulates the separation of the mirrors in an arm, typically by less than $10^{-19}$\,m around 100\,Hz. To measure such small displacements, the optical cavities are held on resonance \cite{Stanley2014} with highly-stabilized laser light in order to both build up laser power and to ensure a linear response to the GWs. Typical ground motion of $10^{-6}$\,m rms requires state of the art seismic isolation to reach the small enough relative mirror motion needed to detect GWs.

The high level of isolation of the mirrors from the ground, $10$ orders of magnitude at $30$\,Hz and an order of magnitude at $150$\,mHz, is achieved through the use of a series of passive and active control techniques \cite{Biscan2018}. The mirrors are suspended as pendula, which, in turn, are mounted on actively isolated platforms that rely on both feed--forward and feed--back control from sensors on the ground and the platform, respectively. Despite such phenomenal levels of isolation, particularly large seismic vibrations nonetheless remain responsible for about 5\% of the unplanned downtime (out of 25-30\%) at both the LIGO Hanford and LIGO Livingston Observatories (LHO and LLO) during their first observing runs, O1 and O2, from 2015 to 2017. Produced by ocean waves, wind, human activity, and, most dramatically, earthquakes, seismic vibrations can cause a loss of resonance of the Fabry-Perot arms, known as a lock loss, when the input motion to the seismic control system exceeds its design capabilities. Excessive actuation to the isolation platforms exacerbates problems arising from imperfect decoupling of degrees of freedom, both on the platform itself as well as further downstream in the optical control of the cavities. After a lock loss, it takes a minimum of 30 minutes to reacquire the locked state of the detector, thus severely hindering its duty cycle. 

Two fundamental types of seismic waves are generated by an earthquake: body and surface waves. Body waves propagate within a body of rock and are subdivided into the faster Primary wave (P-wave), or longitudinal--compressional wave, and the slower Secondary wave (S-wave), or shear wave. 
 
When shallow earthquakes occur (at depths of several km below the surface), coupled P--S waves known as Rayleigh waves are generated and particularly affect GW detectors because their motion is restricted to near the surface of the ground. Rayleigh waves \cite{Bullen1986,Stein2003} have periods of anywhere from 3--60\,s with the majority occurring at 15--20\,s, corresponding to frequencies of about 50--60\,mHz. The surface waves from a large earthquake can be measured around the world. These teleseismic waves impact Advanced LIGO operations when they are larger than about 1\,$\mu$m/s at the observatories, which happens several times per week on average.

This paper focuses on the design and implementation of a novel set of techniques to reduce the detrimental effect of earthquakes on the GW detectors. Section \ref{sec:ISI} describes the active seismic isolation system used for each core mirror and highlights the problem of how the primary ground motion frequency that is amplified by earthquakes, $50$\,mHz, also gets amplified by the seismic control system when in its nominal configuration. Section \ref{sec:EQmode} presents the solution which consists of two independent, yet complementary, modifications to the seismic control system to reduce the input signal at 50\,mHz and its amplification. Section \ref{sec:results} presents the results of implementing the new \emph{EQ mode} at LLO and LHO during observing run 3 (O3). Finally, Section \ref{sec:conclusions} discusses the major accomplishments of this work and provides an outlook for the future.

\section{Advanced LIGO seismic controls}
\label{sec:ISI}
The Advanced LIGO mirrors are suspended from isolation platforms. These platforms fulfil several roles. They provide active positioning from DC to about 30 Hz, active isolation from ground motion from around 100\,mHz to 30 Hz, and passive isolation above a few Hz. The multi-stage mirror suspensions provide additional passive isolation, active damping, and control for the optics.

The actively isolated platforms make up part of the Internal Seismic Isolation (ISI) system, in which a number of sensors on the ground and on the platform are used for feed--forward and feed--back control. Each platform's motion is reduced to $10^{-12}$\,m/$\sqrt{\mbox{Hz}}$ at 10\,Hz, several orders of magnitude below ground motion. Details about the design, operation and performance of the ISI platforms can be found in \cite{Matichard,Abbott2002,Abbott2004}.

\begin{figure}[h!]
\includegraphics[scale=0.475]{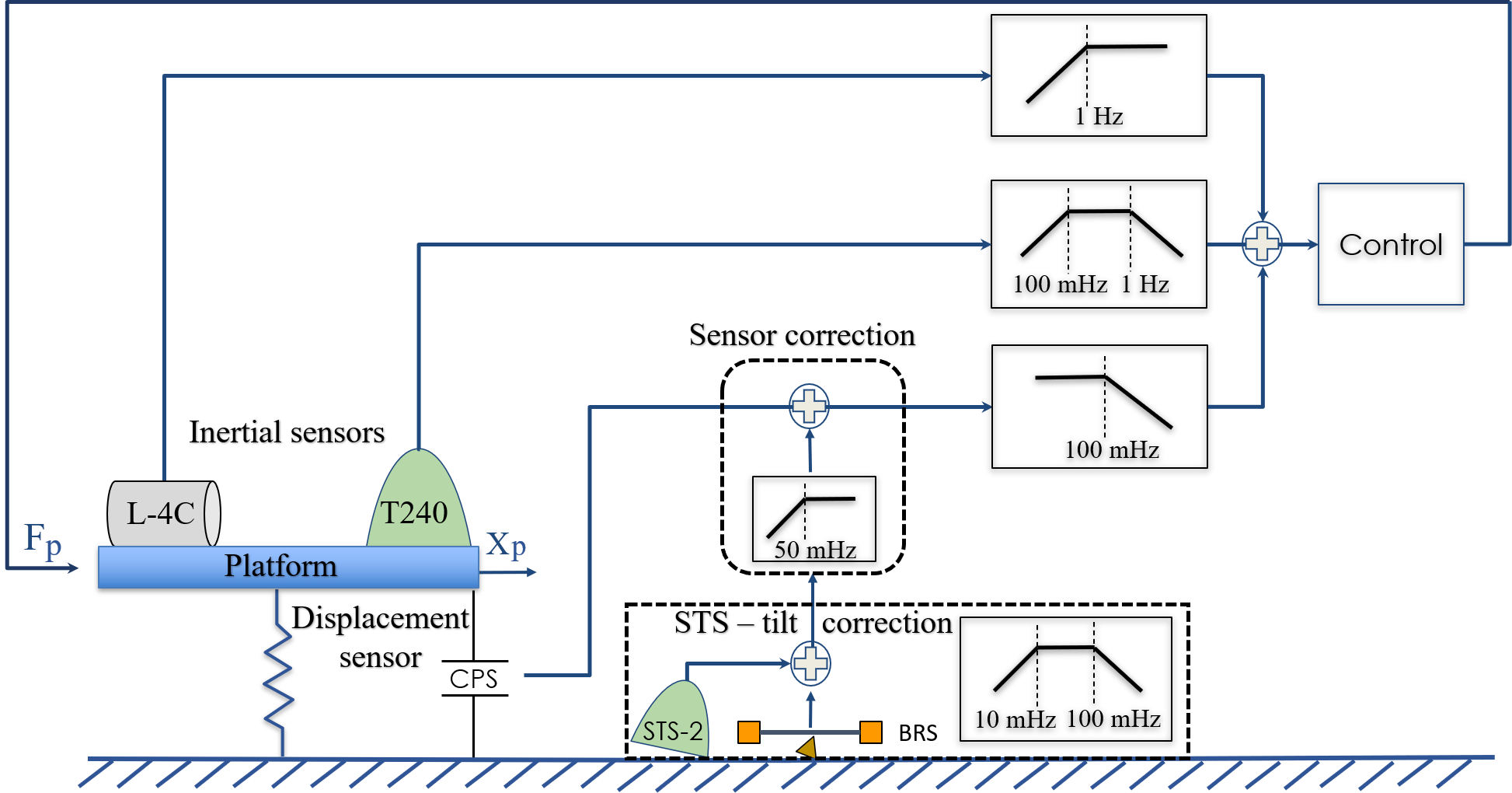}
\caption{Illustration of the BSC--ISI Stage 1 seismic controls. Sensors on the ground and on the ST1 platform are combined to act as a witness of the platform's inertial motion. Control signals are derived and sent to actuators on the platform. Note that multiple sensors (not drawn) of each type are used to provide control in all three Cartesian degrees of freedom.}
\label{figureone}
\end{figure}

Two types of ISI systems are used at the observatories. The HAM-ISIs are single--stage platforms which support the power and signal recycling mirrors, as well as auxiliary optics, and are located in small vacuum chambers called horizontal access modules (HAMs). The BSC-ISIs are two--stage platforms for the beam-splitter and the four mirrors of the arm cavities and are located in large vacuum tanks called basic symmetric chambers (BSCs) \cite{Matichard}. In this paper we will focus on the first stage of the BSC--ISI. It is from this Stage 1 (ST1) platform that a second platform is suspended, from which the mirrors in turn are suspended. 

A schematic of the active seismic control for ST1 of the BSC--ISI is depicted in Figure \ref{figureone}. An absolute inertial measurement of the platform motion, $X_P$, is constructed across a broad band of frequencies from a combination of sensors on the ground and on the platform. The sensed motion is used to generate a feed--back control signal, $F_P$, which is applied to the platform via actuators to reduce the overall motion. 

Information about the platform's motion above $100$\,mHz is provided by two types of inertial sensors located on the platform: the Trillium 240 (T240) \& the Sercel L--4C geophone \cite{seismometers2014,L4C,L4C2}. Information below $100$\,mHz is provided by displacement sensors on the platform, the Microsense Capacitive Position Sensors (CPS), and a tilt-corrected Streckeisen STS-2 seismometer located on the floor near the BSC \cite{seismometers2014,cps}. 

The CPS measures the relative position between the ground and the ST1 platform. In order to use the CPS as an inertial sensor of the platform's motion, information about the ground's absolute motion is needed. A scheme denoted as \emph{sensor correction} (SC), in which the signal from the STS-2 is first high--pass--filtered and then added to the CPS signal turns the CPS into a virtual inertial sensor at frequencies above $50$\,mHz \cite{Lantz2012}. The filtering of the STS-2 signal, using a filter called the \emph{sensor correction filter}, is necessary because at frequencies below $\sim 50$\,mHz, the STS-2 measurements of ground motion are contaminated both by sensor noise and by unresolved tilt of the ground due to the fact that conventional seismometers cannot differentiate between horizontal displacement and ground tilt \cite{Ross2018}. Ground tilt that can be measured is removed from the STS-2 signal through a scheme called \emph{STS--tilt correction}, which uses a beam rotation sensor (BRS) that measures the tilt of the ground near the STS-2 \cite{Ross2018}.

\begin{figure}[h!]
\includegraphics[scale=0.48]{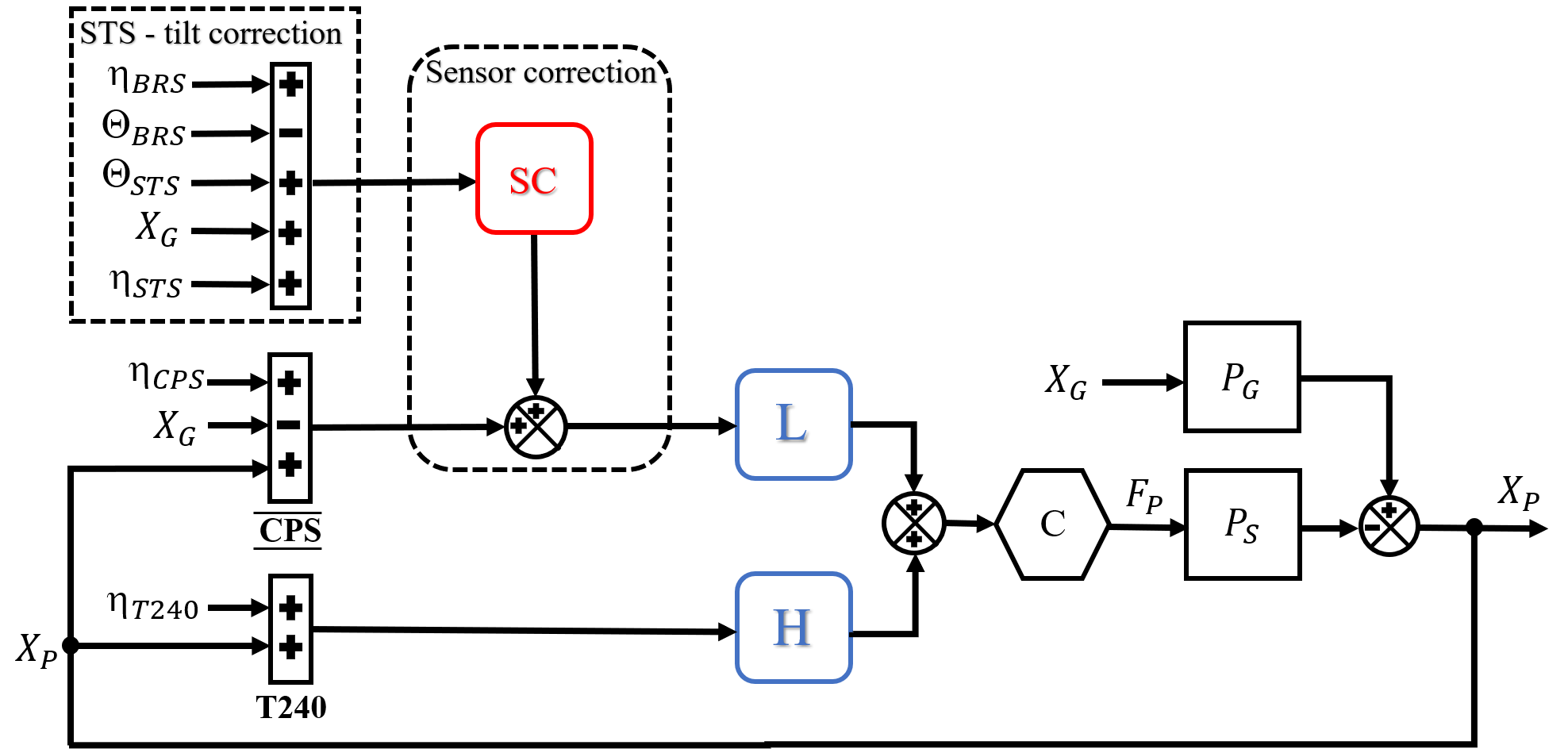}
\caption{Block diagram of the control servo up to $500$\,mHz for one degree of freedom of a BSC--ISI ST1 platform. The L-4C is not depicted because it only contributes to the control above $\sim 1$\,Hz.}
\label{figuretwo}
\end{figure}

A block diagram detailing the design of the nominal control scheme for a BSC--ISI ST1 platform up to 500\,mHz is shown in Figure \ref{figuretwo}. Inputs representing sensor noises ($\mathlarger{\mathlarger{\eta}} _{BRS}$, $\mathlarger{\mathlarger{\eta}} _{STS}$, $\mathlarger{\mathlarger{\eta}} _{T240}$ and $\mathlarger{\mathlarger{\eta}} _{CPS}$) and actual ground ($X_G$) and platform ($X_P$) motion are shown on the left and the resulting inertial motion of the platform on the right. The tilt--corrected ground motion is given by $\widetilde{X}_G = X_{G}+\Theta_{STS}-\Theta_{BRS}+\mathlarger{\mathlarger{\eta}} _{STS}+\mathlarger{\mathlarger{\eta}} _{BRS}$, where $\Theta_{STS}$ and $\Theta_{BRS}$ is the ground tilt sensed by the STS--2 and the BRS correspondingly. These two tilt contributions cancel each other out, therefore $\widetilde{X}_G$ becomes limited by the noise from one of the two instruments, $\mathlarger{\mathlarger{\eta}} _{STS}$ or $\mathlarger{\mathlarger{\eta}} _{BRS}$. The CPS signal by definition is $X_P - X_G + \eta_{CPS}$. The sensor--corrected CPS signal is low--pass--filtered ($L$) and blended with the high--pass--filtered ($H$) inertial sensors. $P_S$ and $P_G$ are the plant transfer functions, describing how the platform reacts to the actuation and ground motion, respectively. The control filter, $C$, is designed such that the open loop gain is unity at $\sim 30$\,Hz, and is very large at low frequencies \cite{Matichard2}.  

When feed--back is engaged, the expected inertial motion of the platform below $\sim 1$\,Hz is thus given by
\begin{equation}
\begin{split}
X_P & = \frac{P_G X_G+(X_G-\widetilde{X}_G \cdot \textup{SC})\cdot L \cdot C \cdot P_S}{1+C\cdot P_S}
-\frac{\mathlarger{\mathlarger{\mathlarger{\eta}}} _{BRS}\cdot \textup{SC}\cdot L\cdot C\cdot P_S}{1+C\cdot P_S}\\&-\frac{(\mathlarger{\mathlarger{\mathlarger{\eta}}} _{CPS}\cdot L+\mathlarger{\mathlarger{\mathlarger{\eta}}} _{T240}\cdot H)\cdot C\cdot P_S}{1+C\cdot P_S},
\end{split}
\label{eq1}
\end{equation}
which can be simplified to
\begin{equation}
\underset{C\cdot P_S\gg1}{X_P} = X_G\cdot(1-\textup{SC})\cdot L-\mathlarger{\mathlarger{\eta}} _{BRS}\cdot \textup{SC}\cdot L-(\mathlarger{\mathlarger{\eta}} _{CPS}\cdot L+\mathlarger{\mathlarger{\eta}} _{T240}\cdot H)
\label{eq2}
\end{equation}
given the very large gain ($C\cdot P_S \gg 1$) at frequencies below 1\,Hz and that $L+H = 1$ by construction. We further assume that $\widetilde{X}_G\approx X_G+\mathlarger{\mathlarger{\eta}} _{BRS}$ because tilt cancels out (i.e. $\Theta_{STS}=\Theta_{BRS})$ and $\mathlarger{\mathlarger{\eta}} _{STS}$  is negligible with respect to $\mathlarger{\mathlarger{\eta}} _{BRS}$ at frequencies below 1\,Hz .

This nominal configuration of the seismic controls is regularly used and enables operation during the majority of the varying environmental conditions, including different microseismic states, wind and small earthquakes ($<1\mu \mbox{m/s}$). To operate through larger earthquakes, this configuration was altered in order to confront the impact of high ground motion on the stability of the detectors.

\section{Earthquake mode}
\label{sec:EQmode}
The goal of a seismic control configuration specifically designed for use during earthquakes is to maintain the optimized performance of the seismic isolation system at frequencies above $100$\,mHz while limiting gain peaking in the $50-60$\,mHz frequency band \cite{Biscan2018}. The result should be a reduction of the fluctuations of the length of each arm, as well as a reduction of the differential motion of the arms with respect to one another. The latter is the degree of freedom known as DARM (differential arm) which dictates the sensitivity of the detector to GWs. \emph{EQ mode} consists of two distinct modifications to the ISI controls: 1) a change to the SC filter; and 2) a change to the input of the SC filter.

\subsection{The sensor correction filter}
The SC filter is a high pass filter designed to maximize the seismic isolation at $100-300$\,mHz, where the ground motion is dominated by the secondary microseism caused by storms in the ocean \cite{Cessaro}, while suppressing any BRS noise or residual tilt below 100\,mHz. The trade off for good isolation above 100\,mHz is amplification of the ground motion at the cut-off frequency of the high pass filter SC. We call gain peaking the maximum amplification in the frequency band where the suppression function (1-SC) is greater than unity. \footnote{The Bode Integral theorem, or `waterbed theorem' shows that, for non--trivial real time control systems, improved performance at some frequencies results in degraded performance at other  frequencies.\cite{Bode}}. 

\begin{figure}[h!]
\begin{subfigure}{0.49\textwidth}
\includegraphics[width=\linewidth]{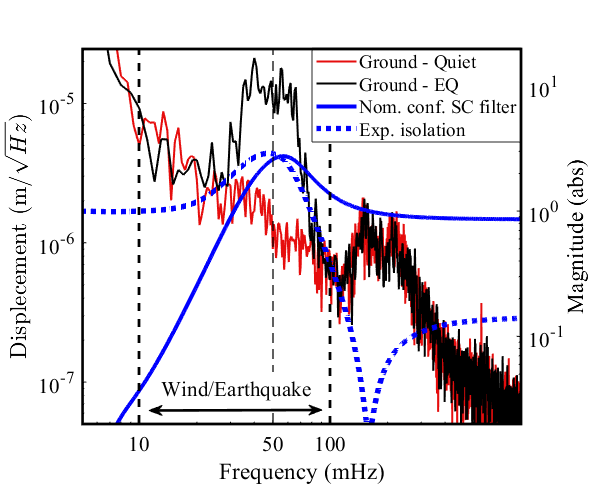}
\caption{} \label{fig:3a}
\end{subfigure}
\hspace*{\fill} 
\begin{subfigure}{0.49\textwidth}
\includegraphics[width=\linewidth]{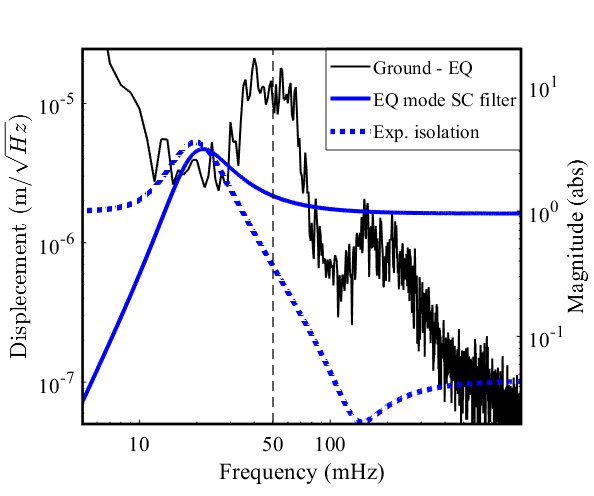}
\caption{} \label{fig:3b}
\end{subfigure}
\hspace*{\fill} 
\caption{(a) Ground motion during an earthquake (black) and quiet times (red) as measured by an STS-2. The nominal SC filter (solid blue) has gain peaking at $50$\,mHz so that $\mbox{SC} \ll 1$ at lower frequencies to ensure that the unresolved tilt and sensor noises are decoupled from the sensor correction. The expected isolation of the ISI platform from the ground, $1-\textup{SC}$, is shown in dashed blue. (b) The SC filter for \emph{EQ mode} has a gain peaking shifted out of the earthquake band down to $20$\,mHz. This is made possible by the decreased false signals at these frequencies during earthquakes.
}
\label{figurethree}
\end{figure}

The nominal SC filter results in gain peaking at 50\,mHz, as shown in \Cref{fig:3a}. \Cref{fig:3a} also plots the ground motion measured by the STS-2 during a quiet time and during a magnitude 5.9 earthquake in Indonesia (2019-10-14 22:23:54 UTC, depth 19.0\,km), clearly showing that the frequency of the gain peaking coincides with the frequency of the peak energy of the earthquake. When the nominal SC filter is used during an earthquake, the amplified excess motion in the $50-60$\,mHz frequency band often results in a lock loss of the interferometer. 

A modification to the SC filter that shifts the gain peaking to lower frequencies is the first of two components that comprise the new \emph{EQ mode} scheme. \Cref{fig:3b} shows the SC filter designed for \emph{EQ mode}, which has a gain peak shifted down to $20$\,mHz. The expected isolation ($1-\textup{SC}$) is also plotted, highlighting how the SC filter for \emph{EQ mode} provides nearly an order of magnitude more isolation at $50$\,mHz compared to the nominal filter, while maintaining the requirements for frequencies above $100$\,mHz.

The shifted gain peaking has the effect of introducing excess motion below $30$\,mHz compared to the nominal SC filter. Tilt motion of the ground and sensor noise of the BRS dominate the motion below $30$\,mHz under normal conditions, therefore restricting our ability to use the modified SC filter all the time. The \emph{EQ mode} SC filter can be used during an earthquake, however, because the character of the ground motion is different than during seismically quiet times. During an earthquake, translational motion of the ground, rather than tilt, dominates at frequencies below $50$\,mHz. 

\subsection{A global control scheme}
\begin{figure}[h!]
\centering
\includegraphics[scale=0.5]{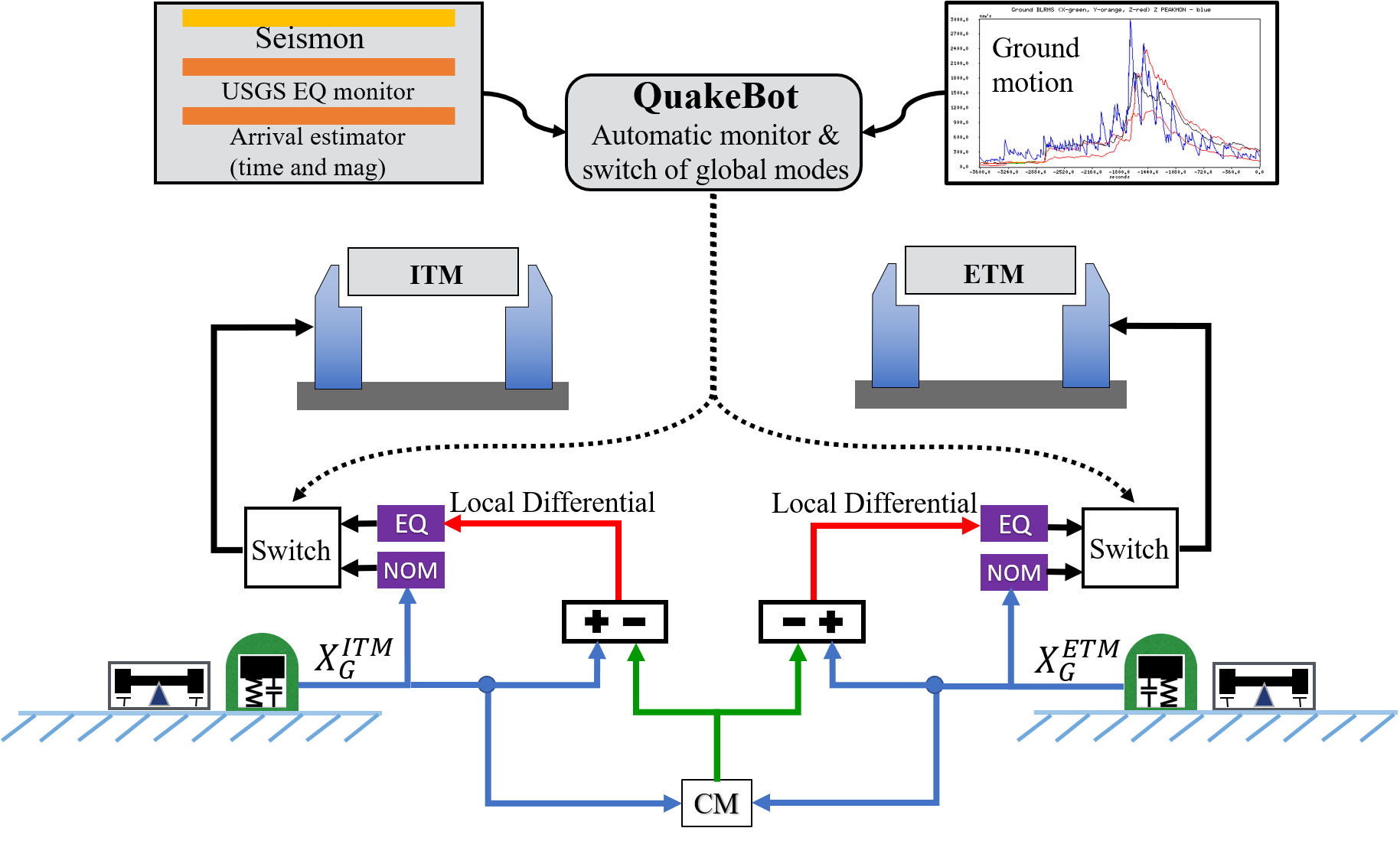}
\caption{Global automatic seismic sensing and control of an arm cavity for one degree of freedom. The local differential is used as an input to the sensor correction of the CPS on the ST1 platform of each station during \emph{EQ mode}.}
\label{figuresix}
\end{figure}

During earthquakes, using the large local ground motion as input to the sensor correction scheme can be problematic. Attempting to keep each ISI platform inertially isolated can saturate the actuators, compromising cavity stability and resulting in a lock loss. The ground motion along the length of an arm during an earthquake is largely coherent, however, and leads to the second modification that comprises \emph{EQ mode}: the removal of the common motion along an arm from the input to the SC filter.
\begin{figure}[h!]
\centering
\includegraphics[scale=0.54]{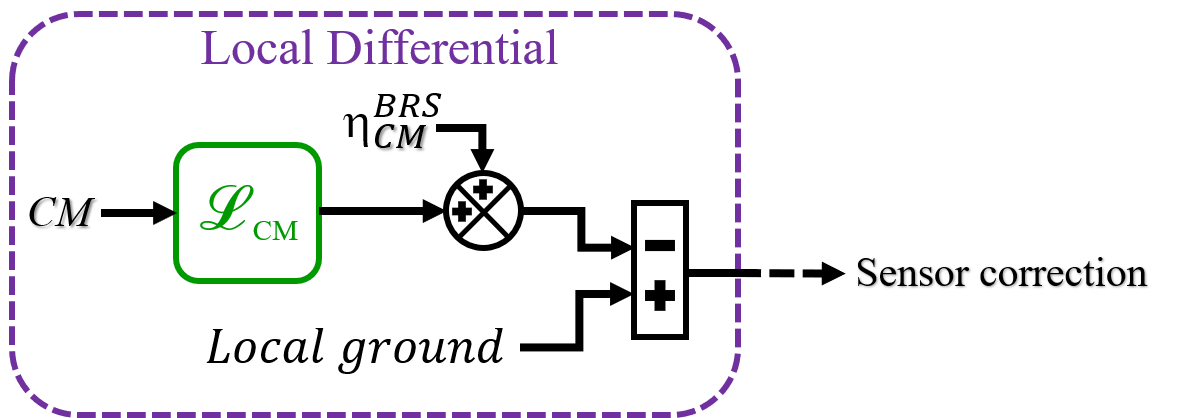}
\caption{Block diagram depicting how the local differential signal is created. During earthquakes, the local differential signal rather than the local ground motion is used as input to the sensor correction (see Figure \ref{figuretwo}). The average of the BRS noise from the two stations of the arm is given by $\mathlarger{\mathlarger{\eta}} ^{BRS}_{CM}$.}
\label{figureseven}
\end{figure}
An analysis of the 134 earthquakes detected at LLO during O3a shows that during earthquakes the common motion of the ground from one end of an arm to the other dominates the differential motion by a factor of $5-6$ in the earthquake band. Defining common--mode as $\mbox{CM} = (X_G^I+X_G^E)/2$, and differential--mode as $\mbox{DM} = (X_G^I-X_G^E)/2$, where $I$ and $E$ represent the corner and end stations, respectively, the ground motion at each station can be represented as a linear combination of DM and CM. Because the differential motion is much smaller than the common motion during an earthquake, removing the common motion component from the input to a platform's control system yields a control signal that falls within the maximum range of the actuators.

We therefore introduced a novel global control scheme which is illustrated in Figure \ref{figuresix}. The STS-2 signals at the corner and end stations are used to calculate the CM motion along the arms. This is in turn subtracted from the local ground motion at each station, thus creating a  \emph{local differential} signal, which is used as the input to the SC filter during earthquakes. By allowing the platforms to sway together, while isolating solely against the differential motion, we create a more stable interferometer.

Figure \ref{figureseven} shows the modification to the sensor correction path for \emph{EQ mode}. The CM of an arm is filtered by a low--pass filter, $\mathscr{L}_{CM}$, which is designed to reject the secondary microseism (100-300\,mHz) while preserving the magnitude and phase of the common mode signal in the earthquake band, below 100\,mHz. The filtered CM signal is then subtracted from the local ground motion thus giving the local differential signal, which serves as input to the sensor correction during an earthquake. Removal of the CM below 100\,mHz is intended to reduce the platform drive when large ground motion occurs. Without the filtering of the CM signal, the secondary microseism would add in quadrature with the local signal, as typically uncorrelated from one end to the other, and would be detrimental for the isolation performance in this band.

\begin{figure}[h!]
\begin{subfigure}{0.49\textwidth}
\includegraphics[width=\linewidth]{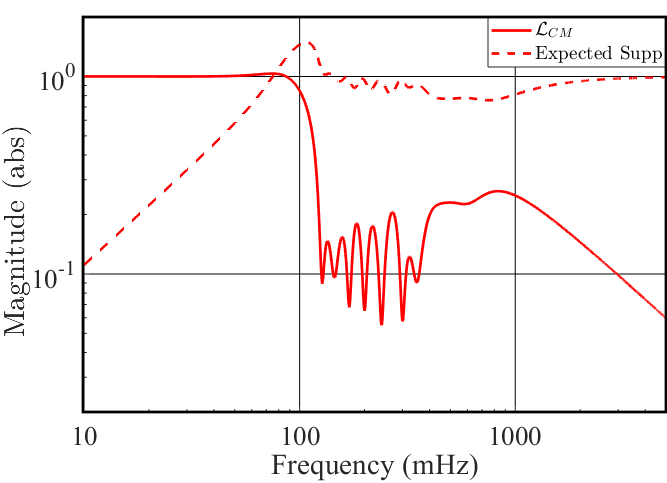}
\caption{} \label{fig:6a}
\end{subfigure}
\hspace*{\fill} 
\begin{subfigure}{0.49\textwidth}
\includegraphics[width=\linewidth]{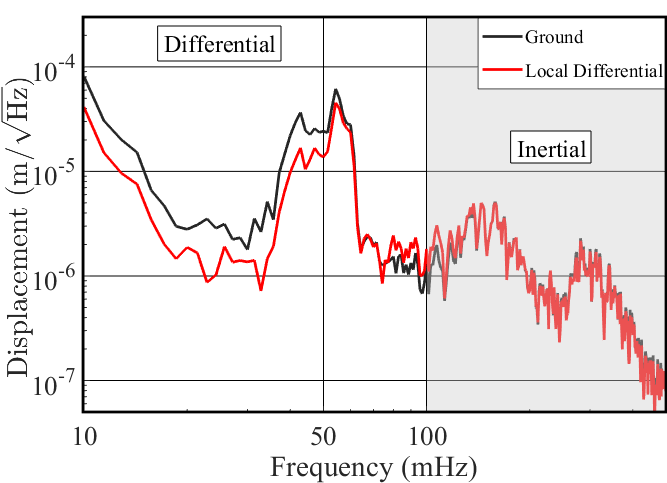}
\caption{} \label{fig:6b}
\end{subfigure}
\hspace*{\fill} 
\caption{(a) The $\mathscr{L}_{CM}$ filter (solid) with the expected CM suppression (dashed). (b) Spectra of ground motion (black) vs. local differential motion (red) created using the CM filter shown in (a).}
\label{figureeight}
\end{figure}

In \Cref{fig:6a} we present $\mathscr{L}_{CM}$ along with the expected reduction of CM motion. \Cref{fig:6b} shows the local ground motion and the resultant local differential motion after $\mathscr{L}_{CM}$ is applied. Here we see one of the impacts of the new \emph{EQ mode} control approach: Up to $50$\,mHz the input to \emph{SC} during an earthquake is reduced by a factor of 2. As seen both in \Cref{fig:6a} and \Cref{fig:6b}, there is a  small increase in noise near 100\,mHz, however the overall reduction in input motion significantly improves the stability of the interferometer.  

\subsection{The EQ mode model}
Our complete model of the ST1 platform motion for the nominal and \emph{EQ mode} configurations is given by:
\begin{subequations}
\begin{align}
\label{eq5a}
X_P^{nom}&=[X_G\cdot(1-\textup{SC}^{nom})-\mathlarger{\mathlarger{\eta}} _{BRS}\cdot \textup{SC}^{nom}+\mathlarger{\mathlarger{\gamma}} X_\perp\cdot \textup{SC}^{nom}]\cdot L\\&-(\mathlarger{\mathlarger{\eta}} _{CPS}\cdot L+\mathlarger{\mathlarger{\eta}} _{T240}\cdot H)\nonumber\\
X_P^{EQ}&=[X_G\cdot(1-\textup{SC}^{EQ})+\textup{CM}\cdot \mathscr{L}_{CM}\cdot \textup{SC}^{EQ}+\mathlarger{\mathlarger{\gamma}} X_\perp\cdot \textup{SC}^{EQ}\label{eq5b}\\
\nonumber&-(\mathlarger{\mathlarger{\eta}} _{BRS}-\mathlarger{\mathlarger{\eta}} ^{BRS}_{CM})\cdot \textup{SC}^{EQ}]\cdot L-(\mathlarger{\mathlarger{\eta}} _{CPS}\cdot L+\mathlarger{\mathlarger{\eta}} _{T240}\cdot H)\nonumber.
\end{align}
\label{eq5}
\end{subequations}

Eq.\,\ref{eq5a} describes the nominal ST1 platform motion and is identical to Eq.\,\ref{eq2}, except for the introduction of an additional linear term, $\mathlarger{\mathlarger{\mathlarger{\gamma}}} X_\perp\cdot \textup{SC}\cdot L$, which describes a residual horizontal motion of the platform that correlates strongly with the platform's vertical motion, $X_\perp$. The importance of the coupling term for producing a realistic model is apparent in Figure \ref{figure12}, though the source of the coupling is not yet understood and remains and active area of investigation. The platform motion when \emph{EQ mode} is engaged is given by Eq.\,\ref{eq5b}. Eq.\,\ref{eq5b} includes the vertical--to--horizontal coupling term in addition to the change in the SC filter and the common--mode subtraction. 

The model proves to be a powerful tool for optimizing the design of \emph{EQ mode} because the transfer function from ST1 to the optics can be approximated as unity below the first resonant mode frequency of the pendula ($400$\,mHz). We can therefore infer how the optical cavities will react in the earthquake band frequencies by simply modeling the behavior of the platforms. 

To analyze the expected and actual performance of \emph{EQ mode}, we focus on the extent to which both the length fluctuations of a single arm cavity and DARM are reduced during an earthquake when \emph{EQ mode} is engaged compared to times when it is not engaged. 

We can estimate the length fluctuations of an arm cavity by calculating the differential motion of the platforms of the arm along the beam direction. This is given by:
\begin{equation}
\Delta_P=\Delta_G\cdot(1-\textup{SC})\cdot L-\mathlarger{\mathlarger{\eta}} _{\Delta BRS}\cdot \textup{SC}\cdot L+\mathlarger{\gamma} _\Delta \Delta_\perp\cdot \textup{SC}\cdot L+O(noises)
\label{eq7}
\end{equation}
where $\Delta=\mathbf{1}-\mathbf{2}$ and $O(noises)$  is the sum of all sensor noises. Importantly, $\Delta_P$ does not include any CM contributions, yet it does rely on the SC filter. We can thus use this model to optimize the \emph{EQ mode} SC filter design to minimize length fluctuations of the arm during an earthquake. 

Likewise, an estimate of DARM at frequencies below $100$\,mHz can be made from the models for platform motion in Eq.\,\ref{eq5}. DARM is defined as (ETMX$_X$-ITMX$_X$)-(ETMY$_Y$-ITMY$_Y$) Where ITM is Input Test Mass and ETM is End Test Mass, thus giving:

\begin{align}
\label{eq8}
\textup{DARM}_{\textup{CPS}}&=(\textup{CPS}_\textup{X}^\textup{ETMX}-\textup{CPS}_\textup{X}^\textup{ITMX})-(\textup{CPS}_\textup{Y}^\textup{ETMY}-\textup{CPS}_\textup{Y}^\textup{ITMY})\\
&=X_G^{\textup{DARM}}[(1-\textup{SC})\cdot L-1]+[\mathlarger{\mathlarger{\gamma}} ^{\textup{DARM}} X_\perp^{\textup{DARM}}-\mathlarger{\mathlarger{\eta}} _{BRS}^{\textup{DARM}}]\cdot \textup{SC}\cdot L.\nonumber.
\end{align}

With these sets of equations, we can predict the motion of a single optic as well as various degrees of freedom of the whole interferometer, allowing us to optimize the design of the seismic isolation system to achieve the best optical performance of the detector during earthquakes.

\subsection{Implementation and automatic switching of \emph{EQ mode}}
Due to the complexity of changing the detector configuration both globally and locally in lock, a change in configuration always has the risk of contaminating the observation data or causing a lock loss. It is therefore essential to switch to \emph{EQ mode} only when a problematic earthquake arrives at the site and not at other unnecessary times. For that, an early alert system named \textit{Seismon}, which uses publicly available data acquired from a global network of seismic observatories compiled by USGS, was developed and installed at the observatories \cite{Coughlin2017, Mukund2019}. Based on a machine learning algorithm, \emph{Seismon} predicts the arrival times and velocities of the P and S body waves and, most importantly, the Rayleigh surface waves of an earthquake with a notification latency of several minutes for earthquakes more than 2000\,km away. Using this information and the current state of the interferometer, a decision can be made about whether to engage \emph{EQ mode}.

Both interferometers are governed by a state machine called \emph{Guardian} \cite{Rollins2016} which consists of automation nodes capable of automatic handling of control changes. We designed a Guardian sub--system (see Figure \ref{figuresix}) which monitors \textit{Seismon} alerts together with the local ground motion in all degrees of freedom to initiate the change in configuration to \emph{EQ mode} when the necessary trigger levels are reached. When the earthquake has passed and the ground motion is below the nominal thresholds, the guardian transfers the seismic platform controls back to the nominal configuration. These changes of state only require a few minutes, as it is possible to switch the filters without having to turn off the seismic isolation loops, and thus keep the interferometer locked. The threshold that we have chosen to trigger \emph{EQ mode} is for earthquakes that produce Rayleigh waves larger than 1$\mu$m/s at LLO and 0.5$\mu$m/s for LHO. Our nominal configuration is capable of handling smaller earthquakes.

When engaging \emph{EQ mode}, we change the seismic control configuration for not only all of the ST1 platforms in the BSC chambers, but also for the single platforms of the HAM chambers. The core optics for the power and signal recycling cavities, crucial components of the design of the GW detectors, are located in both the HAM and BSC chambers. The change to the HAM--ISI controls during \emph{EQ mode} follows the same concepts as the changes for the ST1 platform of the BSC-ISI.

\section{Results}
\label{sec:results}
A month--long break in the third observing run separates O3 into 2 parts, named O3a and O3b. \emph{EQ mode} was first implemented at LHO at the start of O3a and at LLO at the start of O3b. We studied the performance of the isolation platforms at LLO during several earthquakes at the end of O3a and beginning of O3b. We verified that the model for the nominal seismic configuration fits well with the data and then used the model to optimize the design of \emph{EQ mode}. Here, we present the reduction we achieved in local platform motion, single arm motion and DARM when \emph{EQ mode} was engaged during two of the earthquakes studied. This in turn enabled the interferometer to stay locked during the high ground motion. Lastly, we give a statistical analysis of the detectors' performance before and after the implementation of \emph{EQ mode}.

\begin{figure}[ht]
\begin{subfigure}{0.49\textwidth}
\includegraphics[scale=0.56]{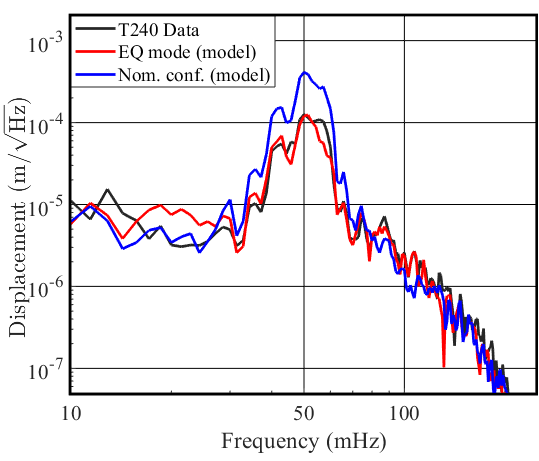}
\caption{} \label{fig:9a}
\end{subfigure}
\hspace*{\fill} 
\begin{subfigure}{0.49\textwidth}
\includegraphics[scale=0.56]{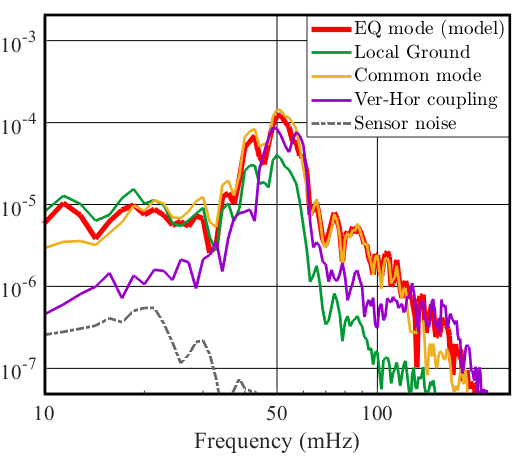}
\caption{} \label{fig:9b}
\end{subfigure}
\hspace*{\fill} 
\caption{(a) Predicted ITMX ST1 platform motion during an earthquake for nominal configuration (blue) and \emph{EQ mode} (red). Data (black) for the platform motion (T240) was acquired when the \emph{EQ mode} was engaged. (b) modeled contributions to the ST1 platform motion that are used to construct the \emph{EQ mode} model in (a) as from Eq.\,\ref{eq5b}. Legend in (b) contains - Local ground: $X_G\cdot(1-\textup{SC}^{EQ})\cdot L$, Common mode: $\textup{CM}\cdot \mathscr{L}_{CM}\cdot \textup{SC}^{EQ}\cdot L$, Ver-Hor coupling: $\mathlarger{\mathlarger{\gamma}} X_\perp\cdot \textup{SC}^{EQ}\cdot L$ and dashed line is the sum of all sensor noise contributions.}
\label{figure9}
\end{figure}

\subsection{Local platform and single arm motion}
At LLO, one of the times \emph{EQ mode} was engaged was during a magnitude 6.3 earthquake, 102\,km WNW of Kirakira, Solomon Islands (2020-01-27 05:02:01 UTC, depth of 21\,km). The interferometer managed to stay locked with ground rms velocities up to 3.9\,$\mu \mbox{m/s}$ for the horizontal direction and 2.5\,$\mu \mbox{m/s}$ for the vertical, a new record for the detectors (with the nominal configuration, the maximum achieved was 2\,$\mu \mbox{m/s}$ for horizontal and 1\,$\mu \mbox{m/s}$ for the vertical direction during an earthquake). Plotted in Figure \ref{figure9} (a) is the modeled and measured single platform horizontal motion of ITMX, $X_P$, for the nominal configuration and \emph{EQ mode} during the Solomon islands earthquake. The \emph{EQ mode} model fits well with the data, thus demonstrating it is a trustworthy tool for further optimizations. At the earthquake peak frequency of 50\,mHz we get a factor 2.6 reduction in both local platform motion and drive, when switching from nominal configuration to \emph{EQ mode}. In Figure \ref{figure9} (b) we show the modeled contributions for \emph {EQ mode} displacement model, as from Eq.\,\ref{eq5b}. 
From Figure \ref{figure9} (b), we see that at 50\,mHz, the biggest contribution to the single platform motion is the common--mode motion and the vertical coupling. Therefore the removal of common-mode was the main reason of the reduction in the single platform drive and motion.

\begin{figure}[h!]
\begin{subfigure}{0.49\textwidth}
\includegraphics[scale=0.565]{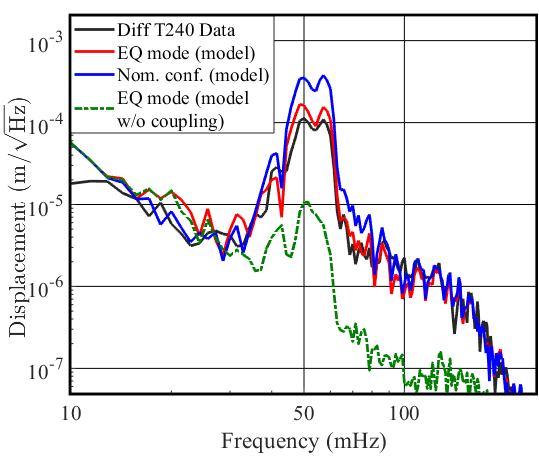}
\caption{} \label{fig:12a}
\end{subfigure}
\hspace*{\fill} 
\begin{subfigure}{0.49\textwidth}
\includegraphics[scale=0.565]{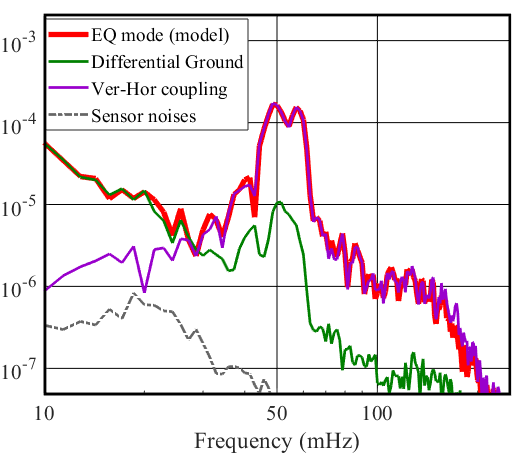}
\caption{} \label{fig:12b}
\end{subfigure}
\hspace*{\fill}
\caption{(a) Predicted differential ST1 platform motion along the X-arm for Nominal (blue) and \emph{EQ mode} (red) during an earthquake. (b) The same as in Figure \ref{figure9} (b), detailing the arm length fluctuations model. Legend in (b) contains - Local ground: $\Delta_G\cdot(1-\textup{SC})\cdot L$, Ver-Hor coupling: $\mathlarger{\gamma} _\Delta \Delta_\perp\cdot \textup{SC}\cdot L$ and dashed line is the sum of all sensor noise contributions.}
\label{figure12}
\end{figure}

In Figure \ref{figure12}(a), we present the arm length fluctuations of the x--arm, as witnessed by the T240 inertial sensors located on each platform of the arm. We also plot our models (Eq.\ref{eq7}) for the nominal configuration and for \emph{EQ mode}. Furthermore, we show the model without the addition of the coupling term, to demonstrate the substantial contribution the coupling term has on the differential motion of the arm. The coupling coefficient per platform, $\gamma$, was empirically determined so that the model optimally fits the data in 30-300\,mHz band. On average, $\gamma\approx0.2$, but does differ between platforms at the different stations. At 50\,mHz we observe a factor 2.5 reduction of arm length fluctuations between the nominal and \emph{EQ mode} configurations. In Figure \ref{figure12} (b) as in Figure \ref{figure9} (b), we show the modeled contributions constructing the \emph{EQ mode} model in (a) as in Eq.\ref{eq7}. We see that the vertical coupling term is the dominant contribution to the single arm differential motion.

\subsection{DARM}

To assess the performance of DARM during earthquakes, when either nominal configuration or \emph{EQ mode} is engaged, we transitioned between these two configurations in the middle of an earthquake to allow close comparison for the same ground conditions and detector optical state (alignment parameters, drift in laser power and gain and optimum optical states change constantly thus affecting performance as well). The data presented is of a 6.1 magnitude earthquake in the Southern East Pacific Rise (2019-12-25 20:20:12 UTC, depth 10.0 km). The maximum rms ground velocities reached 2.3 $\mu$m/s for horizontal motion and 1.3 $\mu$m/s for vertical motion. 
\begin{figure}[ht]
\begin{subfigure}{0.49\textwidth}
\includegraphics[trim={0 0 13.5cm 0},clip,scale=0.565]{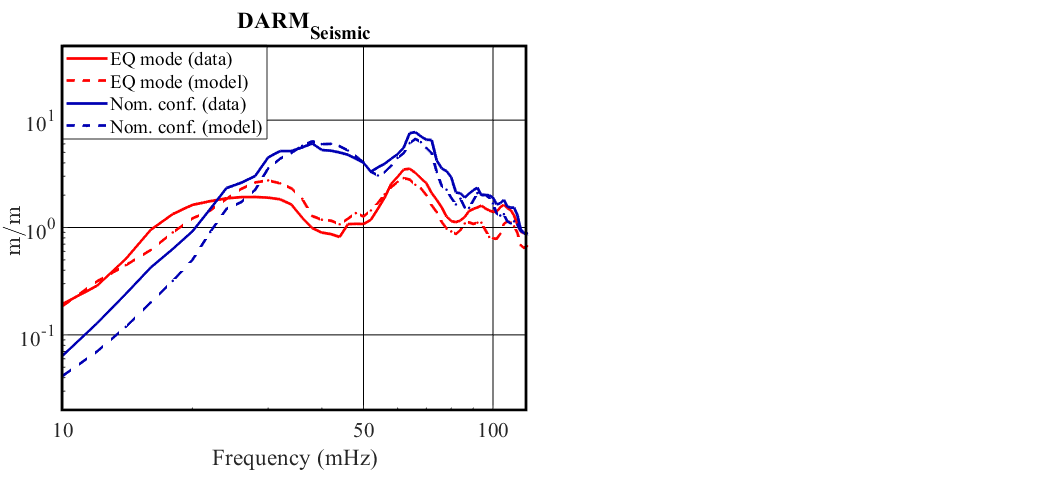}
\caption{} \label{fig:9a}
\end{subfigure}
\hspace*{\fill} 
\begin{subfigure}{0.49\textwidth}
\includegraphics[trim={14cm 0 0 0},clip,scale=0.565]{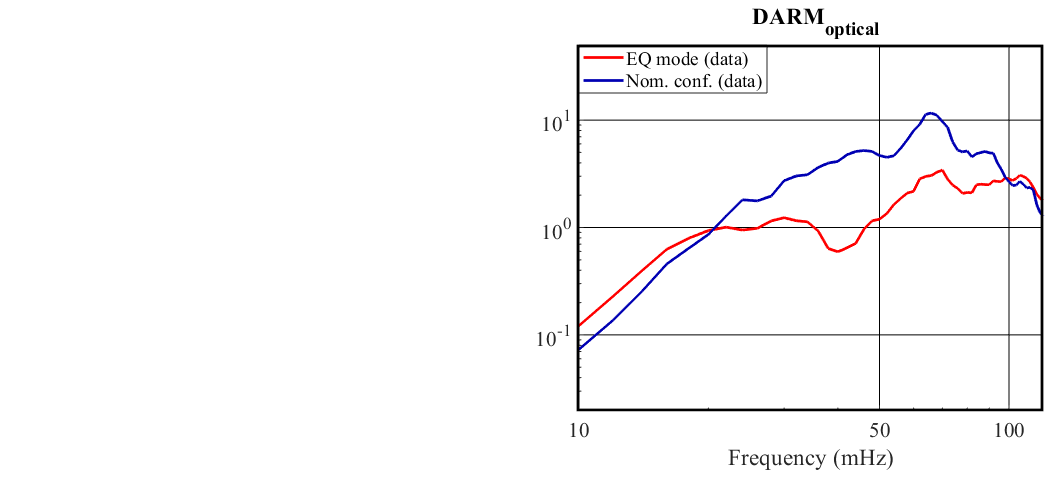}
\caption{} \label{fig:9b}
\end{subfigure}
\hspace*{\fill} 
\caption{(a) Seismic DARM signal. Measured data in solid lines and our model (equation \ref{eq8}) as dashed lines. (b) Optical DARM signal motion. In both sub figures we normalize the plots by the ground motion and blue is nominal configuration while red is \emph{EQ mode}.}
\label{figure14}
\end{figure}

To track the optical signals, we observed the DARM control signal used as input to the ETMX quadruple suspension, which is used to push, or drive, the test mass in order to compensate for the differential arm motion and keep the interferometer locked. Only the top and upper intermediate (UIM) masses had significant and comparable contributions to the total control signal, while the penultimate (PUM) and bottom test mass (TST) were negligible (for LHO, it is the UIM and PUM that contribute to the DARM signal). We calibrated this DARM control signal for sub Hertz frequencies in displacement units \footnote{We used a calibration based on the maximum displacement range at DC per OSEM, per our maximum digital to analog converter range of 18 bits for M0, L1 and L2 \cite{Kissel2014}} .

In \Cref{fig:9a}, we show the seismic DARM signal together with our with our model (Eq. \ref{eq8}). In \Cref{fig:9b}, we plot the optical DARM signal. Both plots are normalized by the ground motion. In this example we get a factor of 4 reduction between the nominal configuration and \emph{EQ mode} at 50\,mHz. From the figure we can see that the optical DARM signal follows the seismic DARM signal, and in turn, our models, in the earthquake band. By this, we show that we can optimize DARM performance during an earthquake by observing the seismic behavior of the platforms and use our models to tune the SC filter and the input to it, and infer from that how the optical cavity will react.

\subsection{Statistical analysis}

\begin{figure}[th]
\centering
\includegraphics[scale=0.7]{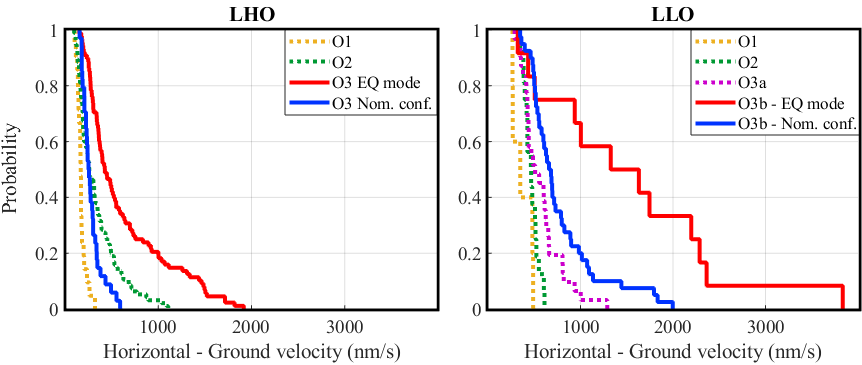}
\includegraphics[scale=0.7]{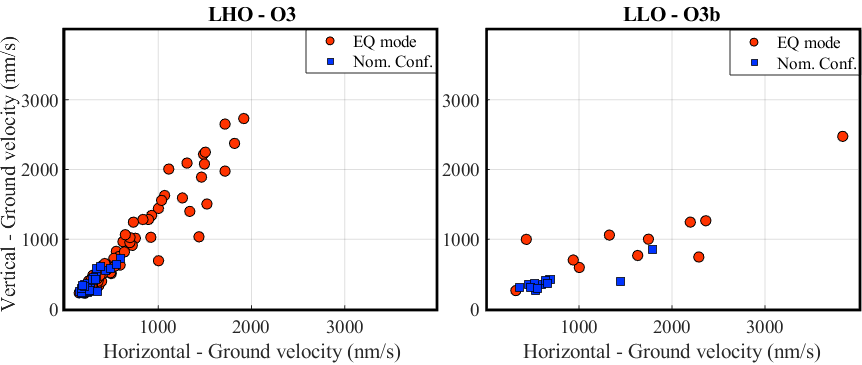}
\caption{Top: The probabilities during each observing run for the LHO and LLO interferometers to stay locked as a function of ground velocity. \emph{EQ mode} was first implemented at LHO in O3a and at LLO in O3b. Bottom: Distribution of ground velocities for the two control modes engaged during earthquakes for which the detectors remained locked during O3 (LHO - left) \& O3b (LLO - right).}
\label{figure13}
\end{figure}

We studied the probability of the LLO and LHO detectors to maintain lock during earthquakes for each observation run starting from O1. The results are presented in Figure \ref{figure13}. We looked at the maximum sustained ground velocities in the beam direction during earthquakes. We included all earthquakes that had yielded a ground velocity rms greater than 200 nm/s in the 30-100\,mHz band limit, sensed by vertical STS-2 at all stations. We then calculated the cumulative distribution functions for each observing run. The two top figures show the clear difference between engaging \emph{EQ mode} versus staying in the nominal configuration during an earthquake. The probability for the interferometer to stay locked during ground velocities above 1000 nm/s for LLO and 500 nm/s for LHO has much increased. For smaller earthquakes, we attribute the improvement to other configuration upgrades carried out. The two bottom figures show the ground velocity distributions for both the horizontal and vertical directions for \emph{EQ mode} and the nominal configuration, during O3 for LHO and O3b for LLO (as in the the top figures). It should be noted that during O3b, the high microseism (100\,mHz--1\,Hz) posed a significant limitation to the performance at both sites, which also limited the use and effectiveness of \emph{EQ mode}.

In Table \ref{tab} we summarize the LLO and LHO  detectors' performance  before (in white) and after (in gray) incorporating \emph{EQ mode}. For LLO, the probability of the detector to stay locked during earthquakes increased significantly from 38\% in O3a to 71\% during O3b when \emph{EQ mode} was introduced. Equivalently for LHO, the increase in probability was from 64\% during O2 to 72\% during O3.

\begin{table}
\centering
\includegraphics[scale=0.7]{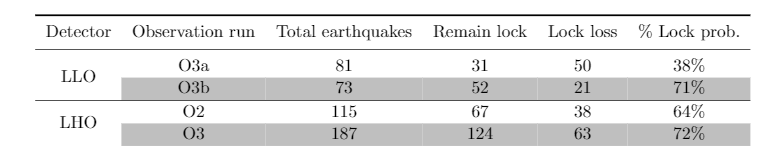}
\caption{LLO \& LHO performance during earthquakes. White rows are data for pre-\emph{EQ mode} (LLO - O3a, LHO - O2 while gray rows are for post implementation of \emph{EQ mode}. The threshold for earthquakes is 200 nm/s on the 30-100\,mHz band limited rms signal of vertical STS-2 at all stations. Total earthquakes relates to all teleseismic events when the detector was in lock just prior to the arrival and managed to stay locked until the end of the event.}
\label{tab}
\end{table}

\begin{figure}[h]
\begin{subfigure}{0.49\textwidth}
\includegraphics[trim={0 0 13.5cm 0},clip,scale=0.565]{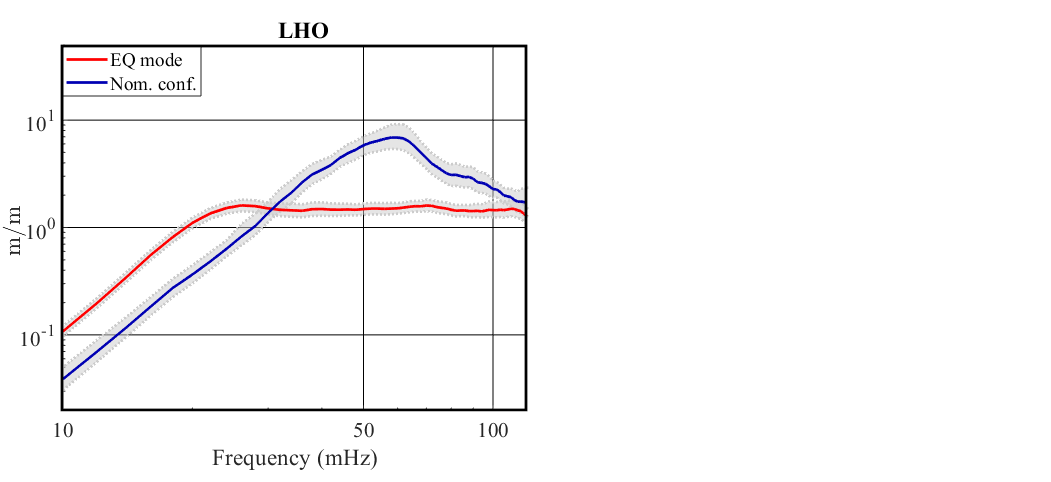}
\caption{} \label{fig:15a}
\end{subfigure}
\hspace*{\fill} 
\begin{subfigure}{0.49\textwidth}
\includegraphics[trim={14cm 0 0 0},clip,scale=0.565]{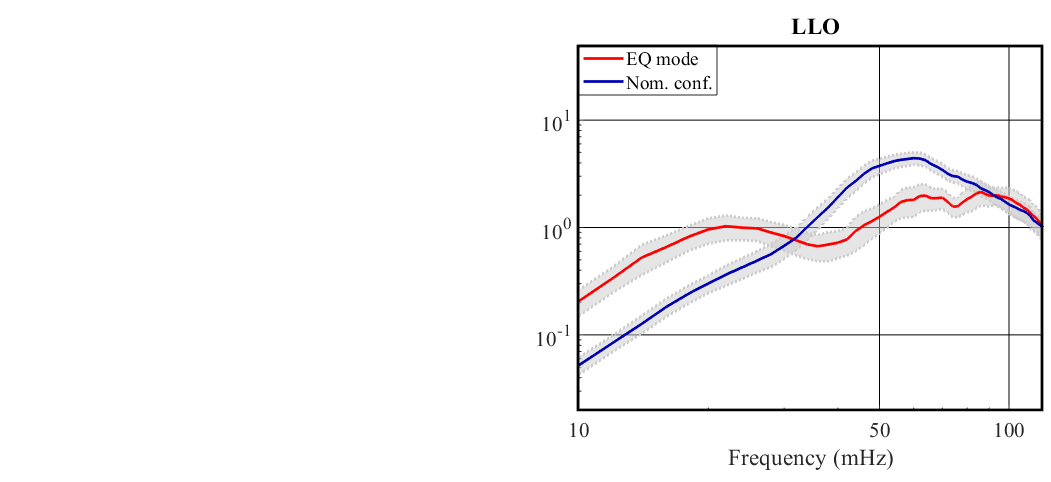}
\caption{} \label{fig:15b}
\end{subfigure}
\hspace*{\fill} 
\caption{Averaged DARM signal normalized by ground motion during earthquakes for which the detector remained locked as given in Table 1 for (a) LHO - O3 and (b) LLO - O3b. Nominal configuration presented in blue and \emph{EQ mode} in red. Gray area is the 95\% confidence interval of the mean for each configuration.}
\label{figure15}
\end{figure}

Lastly, in Figure \ref{figure15} we present a statistical analysis of the LHO and LLO DARM signals (similar to \Cref{fig:9b}) for O3 and O3b correspondingly. We averaged the calibrated DARM signal, normalized by the ground motion, over all the earthquakes for which the interferometers remained locked, as presented in Table \ref{table1}, separated to each seismic configuration (\emph{EQ mode} is in red while nominal configuration is in blue). The shaded gray area per configuration is the 95\% confidence interval of the mean. This allows us to see that at 50\,mHz, we get a reduction factor between nominal configuration and \emph{EQ mode} of 3.4$\pm$0.5 for LLO and 4.2$\pm$0.6 for LHO.

\section{Conclusions}
\label{sec:conclusions}

Earthquakes are one of the major disturbances detrimental to the performance of ground-based GW observatories, limiting their duty cycle and the amount of useful data for GW detection. In this paper we describe the design and implementation of \emph{EQ mode}, a modification to the nominal seismic control scheme, which includes a global control scheme to isolate against the differential motion of the detector arms, along with a change to local sensor correction filters used to decouple the isolation platforms from ground motion. We combined \emph{EQ mode} with a new automated system that receives alerts about incoming earthquakes and switches the detector seismic configuration to handle such events. We improved the ability of the detectors to maintain lock during ground velocities up to 3.9 $\mu$m/s rms, which is a new record for the detectors.

Our seismic models predict the behavior of the local platform, the single arm differential motion and most importantly the DARM optical signal, which allows us to optimize the performance of the detector at frequencies below 1\,Hz during an earthquake. 

Analysis of the averaged optical DARM signal during an earthquakes shows a clear reduction in DARM motion at frequencies below $100$\,mHz upon switching from nominal configuration to \emph{EQ mode}. Furthermore, our model incorporates a newly discovered cross-coupling between the platform vertical and horizontal motions, yielding a realistic model that matches well to the experimental data.

DARM performance in very low frequencies is of great importance since it directly affects the interferometer stability. Moreover, due to different processes such as light scattering from the optics, undesirable sub-hertz signals get up-converted to higher frequencies (30-100\,Hz), and contaminate the data  where the detector is most sensitive to detect gravitational waves. \emph{EQ mode}, as an automated global control scheme, sets a milestone in seismic controls of the Advanced LIGO detectors, significantly reducing DARM motion at sub-Hertz frequencies during earthquakes, thus contributing to sustaining the data quality and stability of the interferometers in extreme environmental conditions.

\section*{Acknowledgments}
The authors thank the LIGO Scientific Collaboration for access to the data and gratefully
acknowledge the support of the United States National Science Foundation (NSF) for the construction
and operation of the LIGO Laboratory and Advanced LIGO as well as the Science
and Technology Facilities Council (STFC) of the United Kingdom, and the Max Planck
Society (MPS) for support of the construction of Advanced LIGO. Additional support for Advanced LIGO was provided by the Australian Research Council. This project was supported by NSF grants: PHY-1708006 and 1608922. ES acknowledge the LSC fellows program for supporting his research at LIGO Livingston Observatory.
LIGO was constructed by the California Institute of Technology and Massachusetts Institute of Technology with funding from the National Science Foundation and operates under cooperative agreement PHY-1764464 . This paper carries LIGO Document Number LIGO-P2000072.

\end{document}